\documentclass[a4paper,11pt]{article}
\usepackage{pos}
\usepackage{sidecap}
\usepackage{setspace}
\usepackage{natbib}
\title{Neutrino and Electromagnetic Signals from Tidal Disruption Events: Bridging the Theory with Observations}
\ShortTitle{Neutrino and Electromagnetic Signals from TDEs}

\author*[a]{Chengchao Yuan}
\author[a]{Walter Winter}
\author[b]{Cecilia Lunardini}
\author[c]{B. Theodore Zhang}
\author[d]{Kohta Murase}
\author[e,f]{Bing Zhang}

\affiliation[a]{Deutsches Elektronen-Synchrotron DESY, 
Platanenallee 6, 15738 Zeuthen, Germany}

\affiliation[b]{Department of Physics, Arizona State University, 450 E. Tyler Mall, Tempe, AZ 85287-1504 USA}
\affiliation[c]{Key Laboratory of Particle Astrophysics, Institute of High Energy Physics, Chinese Academy of Sciences, Beijing 100049, China}

\affiliation[d]{Department of Physics, Department of Astronomy \& Astrophysics, Center for Multimessenger Astrophysics, Institute for Gravitation and the Cosmos, The Pennsylvania State University, University Park, PA 16802, USA; Center for Gravitational Physics and Quantum Information, Yukawa Institute for Theoretical Physics, Kyoto University, Kyoto, Kyoto 606-8502, Japan} 

\affiliation[e]{Department of Physics, University of Hong Kong, Pokfulam Road, Hong Kong, China}

\affiliation[f]{The Nevada Center for Astrophysics, Department of Physics and Astronomy, University of Nevada, Las Vegas, Las Vegas, NV 89154, USA}

\emailAdd{chengchao.yuan@desy.de}

\abstract{This proceeding presents recent results from a joint analysis of time-dependent neutrino and electromagnetic emissions from tidal disruption events (TDEs), using both isotropic wind models and relativistic jets. We discuss constraints from \textit{Fermi} Large Area Telescope (LAT) $\gamma$-ray upper limits on the size of the radiation zone and the maximum energies of accelerated cosmic rays, as well as the resulting neutrino productions from TDEs and candidates, including AT 2019dsg, AT 2019fdr, AT 2019aalc, and AT 2021lwx. {The \textit{Fermi} upper limits correspond to a generic neutrino detection rate of $\lesssim0.01-0.1$ per TDE.} Additionally, we explore multi-wavelength modeling of jetted TDEs with luminous X-ray afterglows — another TDE subclass — by incorporating the dynamics of structured jets with time-dependent energy injection. We also examine the connection between neutrinos and their multi-wavelength counterparts, highlighting implications for future multi-messenger discoveries with IceCube, IceCube-Gen2, KM3NeT, and \textit{Fermi}-LAT.}

\ConferenceLogo{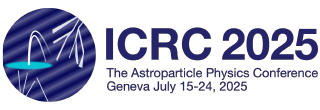}

\FullConference{39th International Cosmic Ray Conference (ICRC2025)\\
 15–24 July 2025\\
Geneva, Switzerland\\}

\begin{document}

\maketitle

\section{Introduction}
Tidal disruption events (TDEs) are luminous optical transients resulting from the tidal destruction of stars that pass within the gravitational influence of a supermassive black hole (SMBH). Approximately half of the disrupted stellar material remains gravitationally bound to the SMBH, and its eventual accretion can produce an electromagnetic flare with a duration ranging from several months to years \citep{1988Natur.333..523R,1989IAUS..136..543P}. Among the approximately one hundred observed TDEs and TDE candidates, four have been identified as being spatially and temporally coincident with astrophysical neutrino events detected by IceCube\footnote{{{These TDEs lie outside the 90\% angular uncertainty regions of neutrino tracks from recent IceCube position reconstructions \cite{Zegarelli:2025vnq}, which weakens the robustness of the correlations. However, our studies of TDEs as potential neutrino sources—particularly the dust echo modeling and the gamma-ray constraints on neutrino production rates—remain valid.}}}. These include one event classified as a TDE with high confidence, AT 2019dsg \citep{2021NatAs...5..510S}, and three TDE candidates, AT 2019fdr \citep{2022PhRvL.128v1101R} and AT 2019aalc \citep{2021arXiv211109391V}, and AT 2021lwx \citep{Yuan:2024foi}, which are associated with neutrino events IC191001A, IC200530A, IC191119A, and IC220405B, respectively. These four TDEs and candidates share several notable characteristics. In particular, they all exhibit high optical and ultraviolet (OUV) luminosities accompanied by bright, delayed infrared (IR) emission, which has been interpreted as dust echoes—reprocessed radiation from the OUV and X-ray bands absorbed and re-emitted at IR wavelengths by surrounding dust \cite{2016MNRAS.458..575L,2016ApJ...828L..14J,2016ApJ...829...19V}, and delayed neutrino detection of $\mathcal O(100~\rm d)$ with respect to the OUV peak. Additionally, Ref. \citep{Jiang:2023kbb} reported a potential association between high-energy neutrinos and two dust-obscured TDE candidates with strong dust echos.

These potential neutrino–TDE associations suggest that TDEs may serve as potential cosmic-ray (CR) accelerators \cite{2009ApJ...693..329F,Zhang:2017hom,2018NatSR...810828B,2023JCAP...11..049P,Plotko:2024gop}, as the astrophysical neutrinos are natural outcomes of hadronic interactions. Many models \citep[see][and references therein]{2020ApJ...902..108M} including relativistic jets, accretion disks, wide-angle outflows/hidden winds, and tidal stream interactions have been proposed as the origin of the non-thermal electromagnetic (EM) and neutrino emission from TDEs, which could potentially explain these TDE-neutrino coincidences. However, the exact radiation sites within TDEs where CR acceleration and neutrino production take place remain under debate. This uncertainty highlights the need for time-dependent, multi-messenger modeling that combines time-domain, multi-wavelength observations with neutrino detections.

Motivated by the delayed neutrino signals and their potential association with infrared echoes, we adopt the quasi-isotropic model proposed by Ref. \cite{2023ApJ...948...42W} and specifically investigate the time-dependent electromagnetic (EM) cascade emissions arising from radiation zones characterized by their typical sizes ($\sim10^{16}-10^{18}~\rm cm$) and the maximum energies of injected protons. In \S \ref{sec:neu_TDEs}, we aim to address several key questions concerning the EM cascade and neutrino emissions from AT 2019dsg, AT 2019fdr, AT 2019aalc, and AT 2021lwx. These include: What observational signatures can be attributed to EM cascades? On what timescales and in which energy bands might these EM cascades manifest? What insights can be gained from X-ray and $\gamma$-ray observations? 

In addition to the quasi-isotropic radiation zone, we also consider the contribution of relativistic jets within the TDE framework, with particular emphasis on four TDEs exhibiting luminous non-thermal X-ray afterglows accompanied by long-lasting radio emission \cite[e.g.,][]{Eftekhari:2024kbb,2023ApJ...955L...6Y}. This TDE subclass includes Swift J164449.3$+$573451 (hereafter, Sw J1644), Swift J2058.4+0516 (hereafter, Sw J2058), Swift J1112.2-8238 (hereafter Sw J1112), and AT 2022cmc. In \S \ref{sec:JettedTDEs}, we examine external reverse shock scenarios to explain the X-ray spectra and light curves of four jetted TDEs, focusing on a fast jet scenario characterized by a Lorentz factor $\Gamma > 10$, while a slower jet component (e.g., $\Gamma \lesssim 5$) may account for the observed radio emission \cite{2023MNRAS.522.4028M, 2024arXiv240611513Y}. We develop a generic, self-consistent model grounded in the TDE accretion history and multi-wavelength observations to describe the jet evolution and time-dependent emissions arising from the reverse shock region. The underlying motivation is that jet deceleration, coupled with an active central engine, jointly governs the reverse shock emission, naturally reproducing the observed X-ray afterglow decay of the form $t^{-\delta}$, where $\delta \sim 5/3$–2.2. We also discuss the detectability of $\gamma$-ray and neutrino signals based on parameter sets that successfully interpret the X-ray observations.

\section{Infrared echoes and neutrinos from TDE isotropic radiation zones}\label{sec:neu_TDEs}

\begin{figure*}
    \centering
           \includegraphics[width=0.42\linewidth]{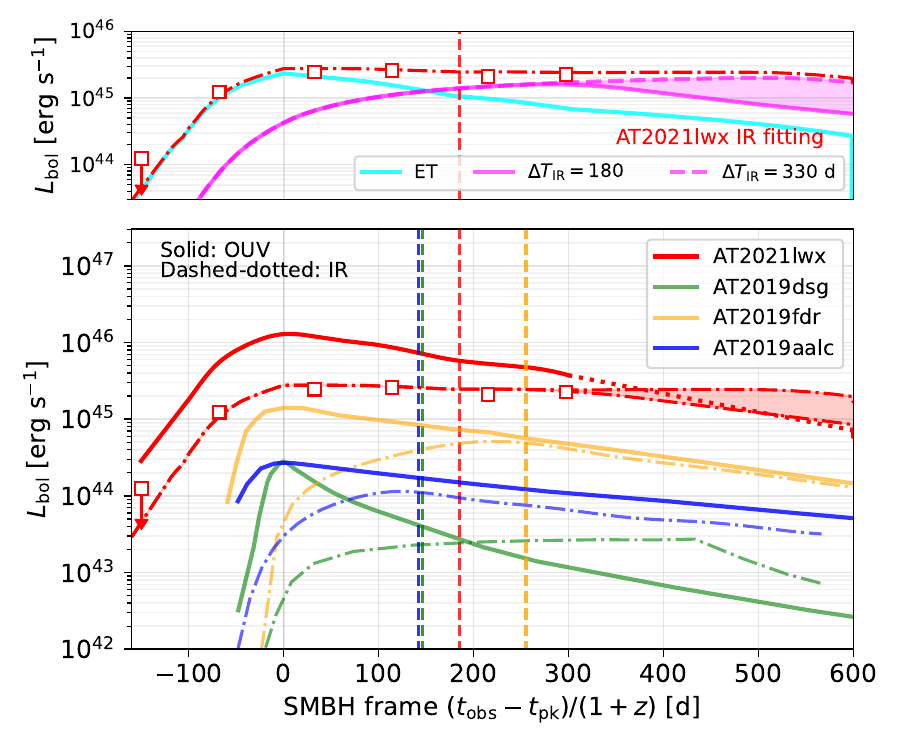}
    \includegraphics[width=0.42\linewidth]{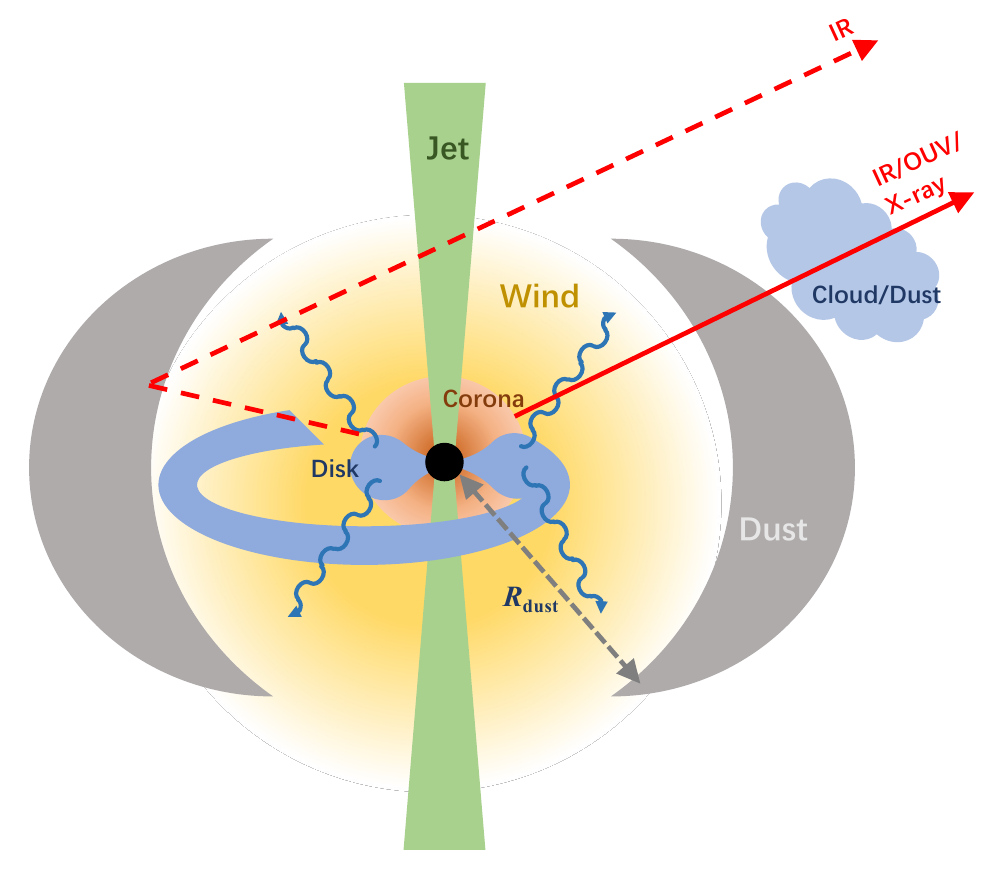}
    \caption{Left panel: {bolometric} OUV (solid curves) and IR (dashed-dotted curves) light curves of AT 2021lwx and the other three neutrino-coincident TDEs in SMBH frame. The vertical dashed lines show the detection times of the corresponding neutrino events. For AT 2021lwx, the IR light curve is interpreted as the superposition of the early-time (cyan) and spherical (magenta) components. Right panel: schematic picture of a TDE with dust echoes. The central SMBH, accretion stream, accretion disk, disk corona, wind, dust torus and potentially a jet are shown. Figures are adapted from Ref. \cite{Yuan:2024foi}.}
    \label{fig:schematic}
\end{figure*}

Four TDEs -- AT 2019dsg, AT 2019fdr, AT 2019aalc, and AT 2021lwx -- exhibit remarkable similarities in their electromagnetic (EM) and neutrino emissions, suggesting a shared physical mechanism. These sources display bright thermal OUV emissions with peak luminosities of $\sim10^{44}-10^{46}~\rm erg~s^{-1}$, powered by super-Eddington accretion onto the SMBH. A key feature of these TDEs is their strong IR radiation as shown in the left panel of Fig. \ref{fig:schematic}, interpreted as a dust echo—reprocessed OUV/X-ray emission from surrounding dust (see the right panel of Fig. \ref{fig:schematic} for the schematic description of the dust torus). The observed IR temperatures $\sim 1000$ K support this interpretation, as they lie below the dust sublimation threshold ($\sim$1800 K). 

The IR light curves show a characteristic delay of $\Delta t_{\rm IR}\sim$100–300 days (in the SMBH rest frame) relative to the OUV peak, corresponding to dust at distances of $R\sim\Delta t_{\rm IR}c/2\simeq10^{17}-10^{18}$ cm. For these four TDEs/candidates, we model the IR luminosity $L_{\rm IR}$ as the convolution of the optical bolometric luminosity $L_{\rm OUV}$ (the solid curves in the left panel of Fig. \ref{fig:schematic}) with a (normalized) time spreading function $f(t)$, which depends on the spatial distribution of the surrounding dust. For AT 2019dsg/fdr/aalc, $L_{\rm IR}$ could be nicely described using a box function $f_{\rm S}(t)=1/(\Delta t_{\rm IR})$ if $0\leq t\leq\Delta t_{\rm IR}$ otherwise $f_{\rm S}(t)=0$, which represents a spherically symmetric dust torus. However, for AT 2021lwx, an additional component $f_{\rm ET}=\delta(t)$ is required to describe the early-time IR light curve produced by the dust close to the SMBH or the iregular/enhanced dust distribution along the line-of-sight \cite{Yuan:2024foi}. Explicitly, the time spread function can be written as $f(t)=\lambda\delta(t)+(1-\lambda)f_{\rm S}$, where $\lambda=0$ and $\lambda\sim0.4$ respectively corresponds to AT 2019dsg/fdr/aalc and AT 2021lwx. The the bolometric IR luminosities obtained from dust echo fit are shown as the dashed-dotted curves in the left panel of Fig. \ref{fig:schematic}.

For the multi-messenger modeling, we focus on a spherical radiation zone of radius $R$, which gives rise to quasi-isotropic emission of both neutrinos and photons. Rather than explicitly modeling the acceleration processes, we parameterize the acceleration zone by the maximum proton energy, $E_{p,\mathrm{max}}$, and assume that the proton injection power constitutes a fixed fraction of the accretion power, expressed as $L_p = \epsilon_p \dot{M} c^2$. Following Ref. \cite{2023ApJ...948...42W}, we adopt a fiducial value of $\epsilon_p = 0.2$ to represent efficient proton injection. The accretion rate is assumed to track the OUV light curve, i.e., $\dot{M} \propto L_{\rm OUV}$, consistent with the canonical $t^{-5/3}$ decline expected from the fallback accretion history. The peak accretion rate is estimated as $\dot{M}(t_{\rm pk}) \sim 100 L_{\rm Edd} / c^2$, ensuring that the total accreted mass does not exceed the physical limit, i.e., $\int \dot{M} dt \lesssim M_\star / 2$, where $L_{\rm Edd} = 1.3 \times 10^{46}{\rm ergs^{-1}}~(M / 10^8 M_\odot)$ is the Eddington luminosity, and $M_\star$ is the mass of the disrupted star. The injection spectrum of accelerated protons is modeled as a power-law with an exponential cutoff, $Q_p \propto E_p^{-2} \exp(-E_p / E_{p,\mathrm{max}})$, which is normalized through the relation $\int E \dot{Q}_p dE_p = L_p / V$, where $E_{p,\rm max}\sim10^8-10^{10}~\rm GeV$ is the proton maximum energy and $V \approx 4\pi R^3 / 3$ represents the volume of the emission region within the dust torus. Within the TDE framework, protons can be accelerated in various regions within TDE systems, including the compact inner jet, the accretion disk or its corona, or in an extended quasi-isotropic wind \citep[e.g.,][]{2020ApJ...902..108M}. The magnetic field strength in these regions is expected to be comparable to that of active galactic nuclei (AGNs), typically in the range of $B \sim 0.1$–1 G.

\begin{figure*}
    \centering
    \includegraphics[width=0.45\textwidth]{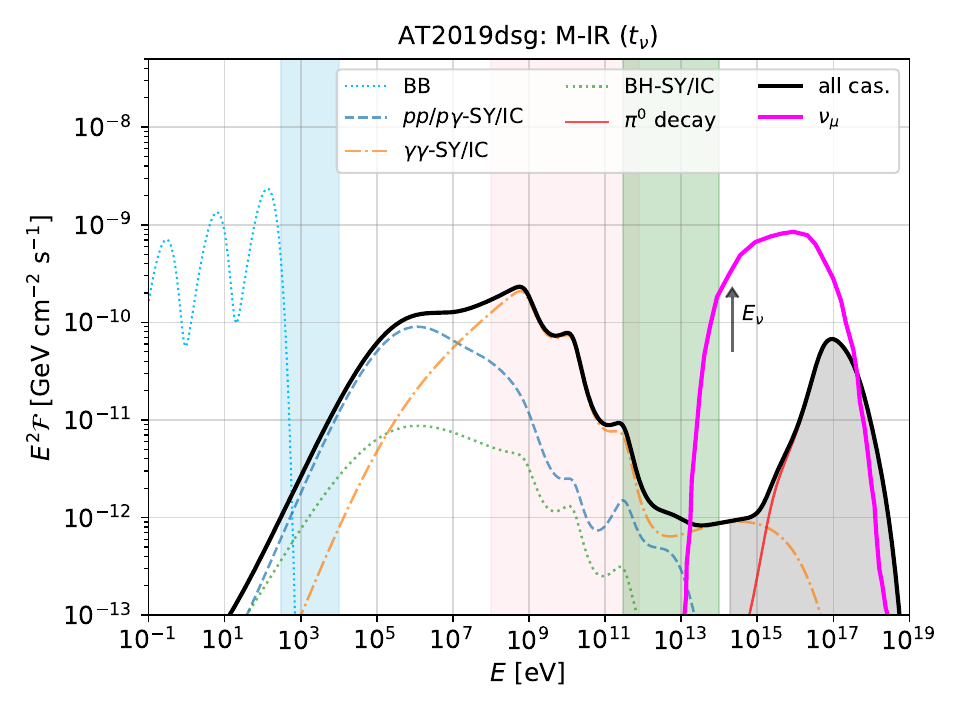}
    \includegraphics[width=0.45\textwidth]{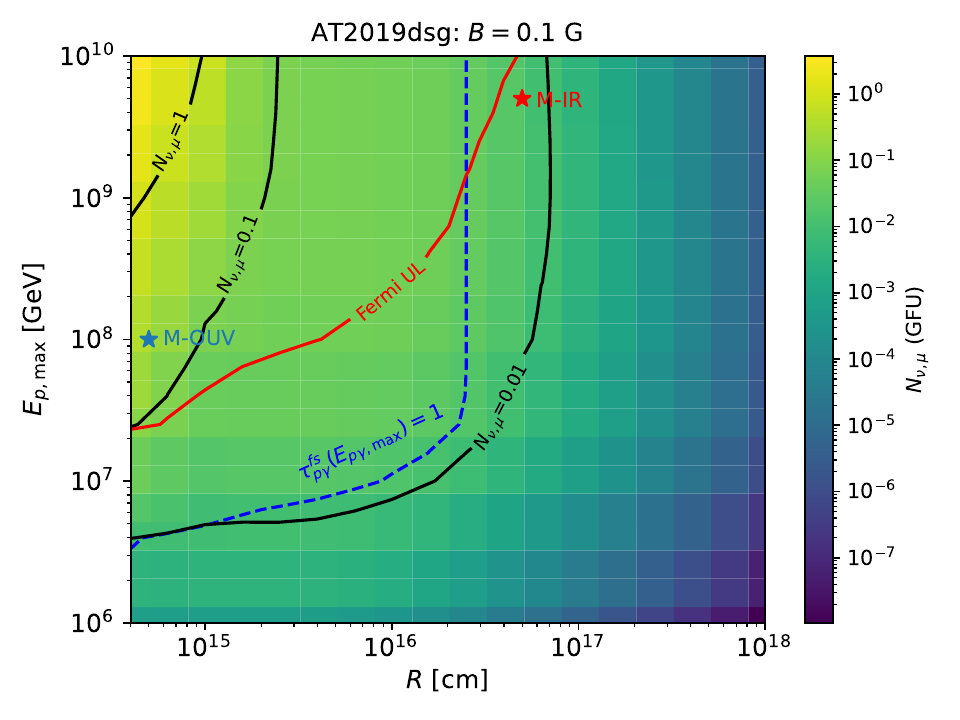}
    \caption{Left panel: EM cascade (black) and neutrino (magenta) spectra at the time of the neutrino observation for AT 2019dsg for the case where IR photons dominates the neutrino production. The blue dotted curve represents the target photon spectra. Right panel: number of expected single-flavor neutrino events as a function of  $R$ and $E_{p,\rm max}$ as given by the color scales. The \emph{Fermi} $\gamma$-ray upper limits on the $R-E_{p,\rm max}$ plane are shown as red curves (the parameters in the upper left corners are excluded), and the regions where the optical thickness  $\tau_{p\gamma}^{\rm fs}(E_{p\gamma,\rm max})=1$ are shown as blue dashed curves (the parameters in the upper left corners are optically thick). Figures are adapted from Ref. \cite{2023ApJ...956...30Y}.}
    \label{fig:Neu_TDEs}
\end{figure*}

While propagating inside the radiation zone where thermal IR/OUV/X-ray photons serve as isotropic target fields, protons undergo energy losses through photomeson ($p\gamma$) and hadronuclear ($pp$) interactions with target photons and wind protons, respectively. These interactions produce neutral ($\pi^0$) and charged ($\pi^\pm$) pions, which decay into neutrinos, $\gamma$-rays, and secondary electrons. The secondary electrons, together with $e^\pm$ pairs generated via $\gamma\gamma$ annihilation and Bethe-Heitler (BH) interactions, initiate EM cascades through synchrotron and inverse Compton radiation.

To obtain the neutrino and EM cascade spectra, we use the AM$^3$ \citep{Klinger:2023zzv} code to numerically solve the coupled time-dependent transport equations for all relevant particle species; see Ref. \cite{2023ApJ...956...30Y} for a detailed description. Taking AT 2019dsg as an example, the left panel of Fig.~\ref{fig:Neu_TDEs} shows the modeled EM cascade and neutrino spectra along with the target photon spectrum based on the dust echo–inferred radius, where IR photons dominate $p\gamma$ interactions. The blue, red, and green shaded regions mark the energy ranges of Swift XRT, \emph{Fermi}-LAT, and HAWC observations, respectively, while the gray-shaded areas indicate suppression due to $\gamma\gamma$ attenuation with EBL. Neutrinos up to PeV energies can be produced, and the delayed neutrino detection is naturally attributed to the combined effects of the $p\gamma$ interaction timescale and the delayed IR emission.

\begin{SCfigure}
    \centering
      \includegraphics[width=0.5\textwidth]{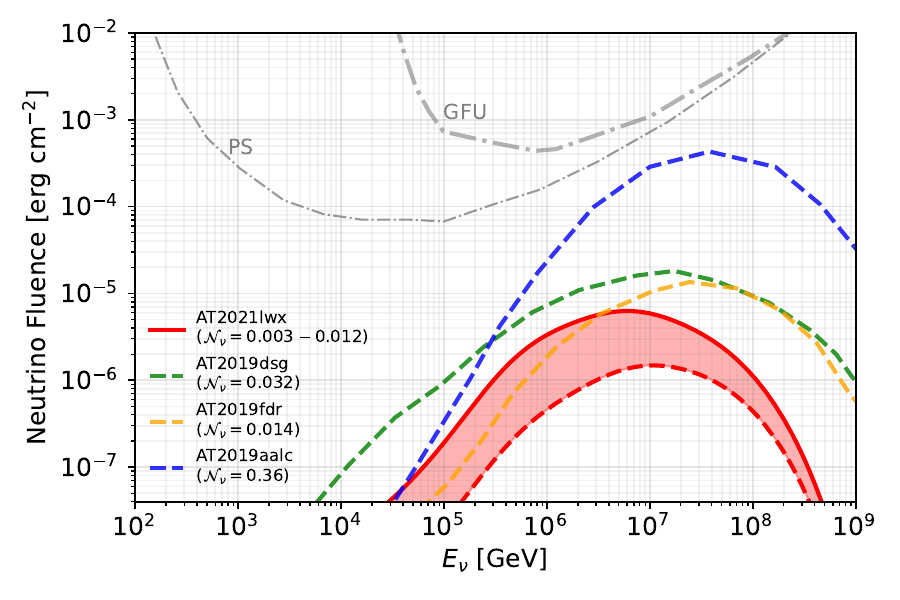}
      \caption{Cumulative single-flavor neutrino fluences at $t_\nu$ for AT 2021lwx (red), AT 2019dsg (green), AT 2019fdr (orange), and AT 2019aalc (blue). The thin and thick dashed-dotted gray curves show the IceCube sensitivities for point-source and GFU searches. {The uncertainties in IR lightcurve interpretation for AT 2021lwx leads to the expected GFU neutrino number in the range $\mathcal N_\nu=3\times10^{-3}-0.012$.} The figure is adapted from Ref. \cite{Yuan:2024foi}.}
    \label{fig:4NeuFluences}
\end{SCfigure}

To explore how the \emph{Fermi}-LAT non-detection constrains the model, we perform a parameter scan on the $R$–$E_{p,\rm max}$ plane. The right panel of Fig.\ref{fig:Neu_TDEs} presents the expected neutrino number as a function of $R$ and $E_{p,\rm max}$ together with the \emph{Fermi} upper limit (red curve). Avoiding the $\gamma$-ray constraint favors an extended radiation zone with optically thin $p\gamma$ interactions and lower $E_{p,\rm max}$, which limits the expected neutrino yield to $\lesssim 0.01-0.1$ per TDE. This conclusion holds for the other TDEs considered. The predicted cumulative neutrino fluences for AT 2019dsg, AT 2019fdr, AT 2019aalc, and AT 2021lwx are shown in Fig.\ref{fig:4NeuFluences}, where the red-shaded regions reflect uncertainties in the IR properties of AT 2021lwx due to the lack of late-time IR data (e.g., the red area in the left panel of Fig. \ref{fig:schematic}).

\section{Jetted TDEs with luminous X-ray afterglows}\label{sec:JettedTDEs}

\begin{figure*}[htp]
    \centering
    \includegraphics[width=0.325\textwidth]{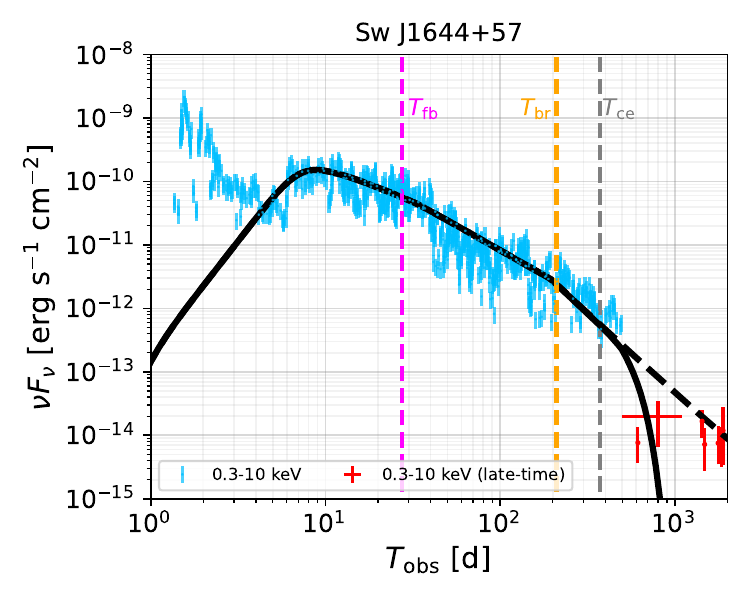}
    \includegraphics[width=0.325\textwidth]{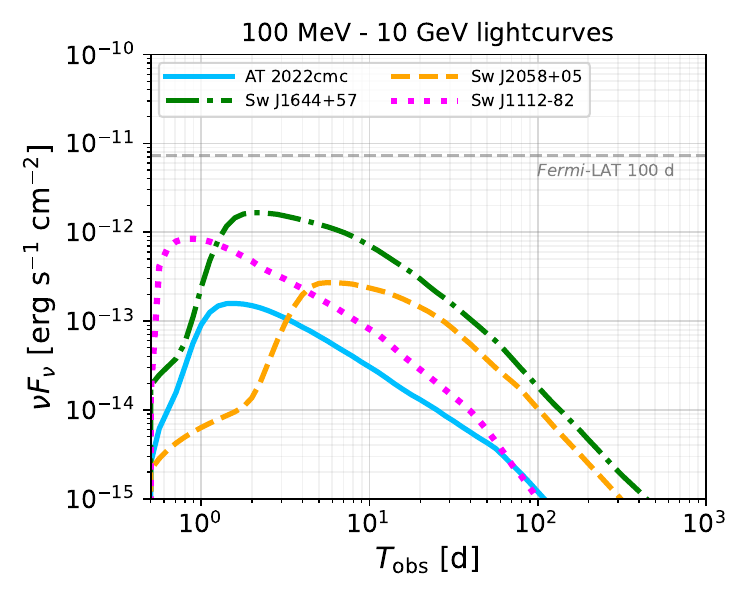}
    \includegraphics[width=0.325\textwidth]{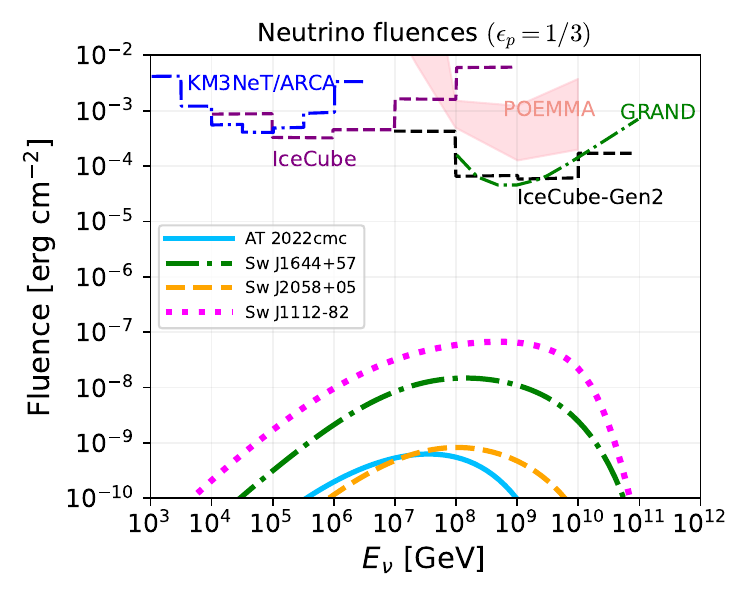}
    \caption{Left panel: Swift 0.3 - 10 keV X-ray light curve fitting for jetted TDE Sw J1644+57. The early-stage and late-time data points are shown as blue and red points. The dashed and solid black curves respectively depict the fitted X-ray light curves before and after accounting for the ceased central engine. The vertical dashed lines indicate the characteristic times including the mass fallback time ($T_{\rm fb}$), jet break time ($T_{\rm br}$), and engine cessation time ($T_{\rm ce}$). Middle panel: model-predicted 100 MeV - 10 GeV $\gamma$-ray light curves for each jetted TDE. The horizontal dashed line indicates the \emph{Fermi}-LAT sensitivity for 100 d observations. {Right panel}: Expected single-flavor neutrino fluences originated from the external FS and RS regions. Optimistic proton acceleration efficiency of $\epsilon_p=1/3$ is used. In both calculations, the parameters obtained from X-ray data fits are applied. The figures are adapted from Ref. \cite{Yuan:2024sxk}.}
    \label{fig:Jetted_TDEs}
\end{figure*}

The second part of this paper investigates four jetted TDEs – AT 2022cmc, Swift J1644, Swift J2058, and Swift J1112 – focusing on their luminous X-ray afterglows and the detectability of the $\gamma$-ray and neutrino emissions. The central hypothesis is that these X-ray emissions originate from the external reverse shock (RS) region of fast relativistic jets, powered by continuous energy injection from the SMBH accretion disk.

The jets are modeled as relativistic outflows with initial Lorentz factors $\Gamma_0 \approx 20-50$, propagating through a circumnuclear medium (CNM) characterized by a density profile $n(R)=n_{\rm ISM}(R/R_{\rm cnm})^{-1.8}$ up to a radius $R_{\rm cnm} \sim 10^{18} \, \mathrm{cm}$, beyond which the density transitions to a uniform interstellar medium (ISM), $n_{\rm ISM}$. The jet power $L_j$ tracks the accretion rate $\dot{M}$, which follows a broken power-law: a shallow decay $\dot{M} \propto t^{-\alpha}$ with $0<\alpha<1$ before the fallback time $t_{\rm fb}$, and a steeper fallback-driven decline $\propto t^{-5/3}$ thereafter. The jet energy budget is expressed as $\mathcal{E}_j = \frac{\eta_j \eta_{\rm acc} M_\star c^2}{2},$ where $\eta_j$ and $\eta_{\rm acc}$ denote the efficiencies of jet launching and accretion, respectively, and $M_\star$ is the disrupted stellar mass. As the jet interacts with the external medium, it drives a FS into the CNM/ISM and a RS into the ejecta. Our model attributes the dominant X-ray emission to synchrotron radiation from non-thermal electrons accelerated at the RS, with a power-law index $s = 2.3$, which could be normalized via the fraction of $L_j$ converted to electrons, e.g., $\epsilon_e$, and the fraction of RS downstream electrons that are accelerated, e.g., $f_e$. The magnetic field strength in the RS region scales as $B_{\rm rs} \sim \sqrt{32\pi\epsilon_B \Gamma_{\rm rel} (\Gamma_{\rm rel}-1)n_0'm_pc^2},$ where $\epsilon_B$ is the magnetic energy fraction, $\Gamma_{\rm rel}$ is the relative Lorentz factor across the shock, and $n_0'$ is the comoving density.

Two main effects shape the X-ray light curves: the jet break occurring when the bulk Lorentz factor decreases to $\Gamma \sim 1/\theta_j$, with typical jet opening angles $\theta_j \approx 0.1$, resulting in a flux drop scaling approximately as $\Gamma^2$; and the cessation of energy injection at a time $t_{\rm ce}$, when the accretion rate falls below the Eddington limit \citep[e.g.,][]{2013ApJ...767..152Z,2014MNRAS.437.2744T}, leading to a rapid decline of the RS emission.

Our model successfully reproduces the observed X-ray light curves and spectra (0.3–10 keV) for all four TDEs. The left panel of Fig. \ref{sec:JettedTDEs} shows the X-ray afterglow fitting for Sw J1644 as an example. The parameters $\alpha=0.65,~\mathcal E_j=3.5\times10^{52}~{\rm erg},~n_{\rm ISM}=1~{\rm cm^{-3}}, \theta_j=0.1,~\Gamma_0=42,~\epsilon_B=\epsilon_e=0.2,~f_e=1.5\times10^{-3},~\eta_j\eta_{\rm acc}=0.02$ are obtained via X-ray light curve and spectral fitting\footnote{See \cite{2023ApJ...955L...6Y,Yuan:2024sxk} for a more detailed description of the model and parameters.}. The early-time rise and shallow decline (e.g.,  $t^{-2.2}$ in Sw J1644) in sources such as Sw J1644 is explained by the RS crossing, while late-time steep decays are consistent with the combined effects of jet breaks and central engine shutdown (after $T_{\rm ce}=(1+z)t_{\rm ce}$ in the observer's frame). Radio emission is attributed mainly to the forward shock of a slower jet component, with Lorentz factors in the range $\Gamma \sim 1-5$. This component dominates the radio band but contributes minimally to the X-ray flux, consistent with multi-wavelength observations.

The predicted synchrotron self-Compton (SSC) emission from the reverse shock peaks at fluxes around $\nu F_{\nu,\rm SSC} \sim 10^{-12} \, \mathrm{erg} \, \mathrm{cm}^{-2} \, \mathrm{s}^{-1}$ in the 100 MeV to 10 GeV range, below the sensitivity threshold of the \emph{Fermi} LAT, as shown in the middle panel of Fig. \ref{fig:Jetted_TDEs}. Klein-Nishina effects and absorption by the EBL further reduce the likelihood of detection at TeV energies.

Assuming that an optimistic fraction $\epsilon_p = 1/3$ of the jet energy powers non-thermal protons, the resulting neutrino fluence in the PeV to EeV range falls approximately two orders of magnitude below the sensitivity of current and upcoming neutrino detectors (see the right panel of Fig. \ref{fig:Jetted_TDEs}, where the cumulative neutrino fluences for the four jetted TDEs are shown) such as IceCube, IceCube-Gen2, KM3NeT/ARCA, GRAND, and POEMMA. Thus, extended on-axis jetted TDEs are unlikely to be prominent neutrino sources unless additional dense target environments, like disk winds, are present to enhance hadronic interactions.

\section{Summary}

Four TDEs (AT 2019dsg, AT 2019fdr, AT 2019aalc, and AT 2021lwx) have been studied as sources of high-energy neutrinos. These TDEs share key observational characteristics, including strong optical-ultraviolet and X-ray emissions, pronounced infrared dust echoes attributed to reprocessing by a dust torus, and neutrino detections delayed by roughly 150 to 300 days relative to the OUV peak in the SMBH rest frame. Time-dependent modeling of neutrino and EM cascade emissions within the dust radius consistently reproduces the delayed neutrino detections. These findings support a common physical scenario where dust echo photons serve as targets for hadronic interactions, producing delayed neutrino emission. The neutrino energy could reach the ultra-high-energy range ($\gtrsim10$ PeV) and predicted neutrino event rates are low, e.g., $\lesssim 0.01-0.1$ per TDE — consistent with the large distances of the sources and the current non-detection in gamma rays by \emph{Fermi}-LAT, which may interpret the reduced likelihood of identifying neutrino–TDE correlations with improved IceCube track reconstructions \cite{Zegarelli:2025vnq}. Correlation studies using the refined angular uncertainties of neutrino tracks are needed for the TDE catalog. 

Separately, the X-ray afterglows of jetted TDEs (Sw J1644, Sw J2058, Sw J1112, and AT 2022cmc) can be comprehensively explained by external reverse shocks in fast relativistic jets ($\Gamma_0 > 10$), while their radio emissions are consistent with forward shocks from slower jets ($\Gamma_0 \lesssim 5$). As relativistic jets with initial Lorentz factors of about 25 to 40 propagate through a circumnuclear medium transitioning to the interstellar medium, the reverse shock energizes electrons that emit synchrotron radiation, shaping the observed X-ray light curves and spectra. Sharp late-time X-ray declines arise naturally from a combination of jet break effects and the cessation of the central engine once the accretion rate falls below the Eddington limit. The model predicts sub-threshold $\gamma$-ray fluxes and neutrino fluences well below current detection sensitivities, consistent with observations. Future improvements incorporating jet precession and internal dissipation processes are expected to better reproduce the detailed variability patterns observed in these jetted TDEs.

This work unifies TDE observations by linking neutrino emissions, EM signals, and jet dynamics into a coherent framework, explaining diverse features across messengers and providing a comprehensive physical picture to guide future multi-messenger studies of TDEs.

{\setstretch{0.8}
\bibliographystyle{JHEP}
\bibliography{ref}}

\end{document}